\documentclass[acmsmall,screen]{acmart}

\AtBeginDocument{%
  \providecommand\BibTeX{{%
    \normalfont B\kern-0.5em{\scshape i\kern-0.25em b}\kern-0.8em\TeX}}}

\acmConference[HATRA '20]{Human Aspects of Types and Reasoning Assistants}{November 18-19, 2020}{Chicago, IL}

\usepackage{enumitem}
\usepackage[frozencache=true,cachedir=minted-cache]{minted}
\usepackage{fancyvrb}

\setminted[rust]{linenos,xleftmargin=15pt}
\setminted[cpp]{linenos,xleftmargin=15pt}

\begin{document}

\title{The Usability of Ownership}

\author{Will Crichton}
\email{wcrichto@cs.stanford.edu}
\affiliation{%
  \institution{Stanford University}
}

\renewcommand{\shortauthors}{Will Crichton}

\begin{abstract}
  Ownership is the concept of tracking aliases and mutations to data, useful for both memory safety and system design. The Rust programming language implements ownership via the borrow checker, a static analyzer that extends the core type system. The borrow checker is a notorious learning barrier for new Rust users. In this paper, I focus on the gap between understanding ownership in theory versus its implementation in the borrow checker. As a sound and incomplete analysis, compiler errors may arise from either ownership-unsound behavior or limitations of the analyzer. Understanding this distinction is essential for fixing ownership errors. But how are users actually supposed to make the correct inference? Drawing on my experience with using and teaching Rust, I explore the many challenges in interpreting and responding to ownership errors. I also suggest educational and automated interventions that could improve the usability of ownership.
\end{abstract}




\newcommand{\rusti}[1]{\mintinline{rust}{#1}}


\maketitle


\section{Introduction}


\noindent Ownership is nominally quite simple to explain. It has three basic rules:
\begin{enumerate}
    \item All values have exactly one owner.
    \item A reference to a value cannot outlive the owner.
    \item A value can have one mutable reference or many immutable references.
\end{enumerate}

\noindent Statically checking these three rules ensures that deallocated memory can never be accessed at runtime. It even rules out unsynchronized concurrent mutation (data races). However, like most interesting program properties, checking whether an arbitrary well-typed Rust program satisfies ownership is undecidable. Instead, Rust uses a sound and incomplete algorithm to verify ownership in a subset of well-typed Rust programs, inspired by prior work in region-based memory management\,\cite{tofte1997region, jim2002cyclone}.

The purpose of this paper is to explore the human factors of ownership-oriented programming in the face of incompleteness. How does a programmer learn to deal with the borrow checker? How can they build an accurate mental model of what the type system will accept and reject? And how can they transfer knowledge from other languages into Rust?

\section{A Litany of Errors}

To get a sense of the problem, let's say a programmer is learning how to use the vector data type. We start with a constructor:

\begin{minted}{rust}
let mut v = vec![1, 2, 3];
\end{minted}

\noindent Under Rust 1.47 (stable as of November 2020), if we try to compile this code:

\begin{minted}{rust}
let x = &v[0];     // Get an immutable pointer to first vector element
v.push(1);         // Attempt to mutate vector
assert!(*x == 1);  // Read from immutable pointer after mutation
\end{minted}

\noindent Then the Rust compiler will raise the following error:

\begin{verbatim}
error[E0502]: cannot borrow `v` as mutable because it is also borrowed as immutable
1 |   let x = &v[0];
  |            - immutable borrow occurs here
2 |   v.push(1);
  |   ^^^^^^^^^ mutable borrow occurs here
3 |   assert!(*x == 1);
  |           -- immutable borrow later used here
\end{verbatim}

\noindent This catch is an objective win --- we avoided an iterator-invalidation scenario where the \rusti{push} reallocates the vector, leaving \rusti{x} pointing to freed memory. Here's a murkier example:

\begin{minted}{rust}
let mut x = &mut v[0]; // Get a mutable pointer to first element
let mut y = &mut v[0]; // Get another mutable pointer to first element
*y += *x;              // Attempt to dereference x and y
\end{minted}

\noindent The compiler will against raise an error:

\begin{verbatim}
error[E0499]: cannot borrow `v` as mutable more than once at a time
1 |   let mut x = &mut v[0];
  |                    - first mutable borrow occurs here
2 |   let mut y = &mut v[0];
  |                    ^ second mutable borrow occurs here
3 |   *y += *x;
  |         -- first borrow later used here
\end{verbatim}

\noindent This program is \textit{technically} memory-safe, but that's contingent on our use of integer data types. If the vector contained vector elements (\rusti{Vec<Vec<i32>>}) then we would have two mutable pointers to the same vector, which could lead to an iterator invalidation scenario like above. It makes sense that this behavior should be generally disallowed. Now let's change the indices to refer to different elements.

\begin{minted}{rust}
let mut x = &mut v[0];
let mut y = &mut v[1]; // Now a mutable pointer to a different element
*y += *x;
\end{minted}

\noindent The compiler will raise the exact same error as before. However, this program is provably safe regardless of data type. Because a vector owns its elements, \rusti{v[0]} and \rusti{v[1]} are guaranteed not to alias. Yet, Rust will reject the program due to incompleteness: it conservatively assumes any indexing operation could refer to any element. The only way to implement this pattern in Rust is to use raw pointers in an \rusti{unsafe} block, or to use the specific stdlib function \rusti{Vec::split_at_mut} (assuming the programmer is aware of it).


In some cases, statements that look virtually identical at the type-level can be treated differently. For example:

\begin{minted}{rust}
    v.insert(0, v[0]);   // Read v[0] and insert it as the 0-th element
    v.get_mut(v[0]);     // Read v[0] and get a mutable pointer to that index
\end{minted}

\noindent Both of these statements use a method that implicitly take a mutable reference to \rusti{v} (i.e. \rusti{&mut self}). However, reading \rusti{v[0]} constitutes taking an immutable reference to \rusti{v}. Today on Rust 1.45, only the second line fails to compile with the error:

\begin{Verbatim}[samepage=true]
error[E0502]: cannot borrow `v` as immutable because it is also borrowed as mutable
1 |   v.get_mut(v[0]);
  |   - ------- ^ immutable borrow occurs here
  |   | |
  |   | mutable borrow later used by call
  |   mutable borrow occurs here
\end{Verbatim}

\noindent About two years ago on Rust 1.30, \textit{both} of these lines would fail to compile. This disparity is due to Rust's new "non-lexical lifetimes" (NLL) algorithm within the borrow checker. NLL increased the overall completeness of the type system, but not always in easily predictable ways. Here, the difference is that \rusti{get_mut} is actually defined on slices (which vectors can implicitly convert into) while \rusti{insert} is defined only on vectors. For whatever reason, that distinction is meaningful enough to cause one line to error and the other to not.

Here's another example of two semantically equivalent statements:

\begin{minted}{rust}
// Two examples of: get first element, pushing 1 if not exists
match v.get(0) { Some(x) => x, None => { v.push(1); &v[0] } };
v.get(0).unwrap_or_else(|| { v.push(1); &v[0] });
\end{minted}

\noindent Again, both lines fail to compile on Rust 1.30 (without NLL), and only the second line fails to compile on Rust 1.44 (with NLL). The reason being that the usage of a closure in the second line restricts the borrow checker to assuming a longer lifetime for \rusti{&v[0]} than without a closure in the first line.

\section{The curse of incompleteness}

The examples show that building a mental model of how ownership \textit{actually} works in Rust is hard.
In my experience, these scenarios are all realistic problems that a Rust programmer may encounter. They explain why the borrow checker is consistently a top reason why developers struggle with Rust or give up on it entirely\,\cite{rustsurvey}. Research on the usability of ownership has similarly found that programmers ``were not prepared to use a type system to address an [ownership] bug that they thought of in a dynamic way''\,\cite{coblenz2019usability}.

A core usability problem is the gap between the simplicity of the goal (three ownership \& borrowing rules) and the complexity of the implementation (borrow checker). I can teach the three rules in a single lecture to a room of undergrads. But the vagaries of the borrow checker still trip me up every time I use Rust! When faced with a borrow checker error, I ask the following questions in order:


\begin{enumerate}
    \item Does my code actually violate the ownership rules?
    \item Can I restructure my code to satisfy the borrow checker?
    \item Is there a function that can bypass the borrow checker?
    \item Do I need to write that unsafe code myself?
\end{enumerate}

\noindent I'd like to talk about each question in greater detail: what makes this question hard to answer, and what resources and pedagogies can help?

\subsection{Does my code actually violate the ownership rules?}

The first challenge is to determine whether an error is caused by unsoundness in the program or incompleteness in the borrow checker. Before all else, a programmer must be aware that this distinction exists. Learning resources about the borrow checker rarely emphasize this fact --- the official Rust Book\,\cite{rustbook}, for example, just teaches the high-level ownership/borrowing rules.

Consider a Rust novice with C++ experience trying to transfer patterns from one language to another. To read a string from a vector, they can write a valid C++ program:

\begin{minted}{cpp}
vector<string> v = {"a", "b", "c"};
string s = v[0];
cout << s << endl;
\end{minted}

\noindent Then, they try a simple rewriting into Rust:

\begin{minted}{rust}
// .into() syntax converts &'static str pointer to owned String
let v: Vec<String> = vec!["a".into(), "b".into(), "c".into()];
let s = v[0];
println!("{}", s);
\end{minted}

\noindent This code fails to compile with the error:

\begin{verbatim}
error[E0507]: cannot move out of index of `std::vec::Vec<std::string::String>`
2 |   let s = v[0];
  |           ^^^^
  |           move occurs because value has type `std::string::String`,
  |               which does not implement the `Copy` trait
\end{verbatim}

\noindent The subtle difference between the two programs is that line 2 of the C++ program implicitly invokes the string's copy constructor, causing a heap allocation. Rust only implicitly copies types that implement the \rusti{Copy} trait (e.g. integers, pointers) that are bit-for-bit copyable without invoking the heap allocator. This distinction is fundamental to ownership --- a Rust programmer must unlearn implicit copying patterns acceptable in other languages. Here, they either need to take a reference (\rusti{&v[0]}) or deep-copy the data (\rusti{v[0].clone()}).

However, the programmer later tries to mutably index disjoint parts of the array, e.g. \\ \rusti{&mut v[0]; &mut v[1]} like above. This pattern is also acceptable in C++, but nominally disallowed in Rust. The Rust novice may incorrectly assume that this limitation is fundamental to sound ownership, not understanding that the borrow checker's rejection is purely a matter of incompleteness. Resources for teaching Rust should consider and explicit focus on the sound vs. incomplete distinction, and giving examples of when borrow checker errors fall into one category or another.


\subsection{Can I restructure my code to satisfy the borrow checker?}

If the programmer has decided their code is actually sound, the next step is to move the program out of the incompleteness zone by small semantics-preserving transformations. In this step lies a dark, perverse bag of tricks that all Rust programmers must learn. For example, consider this problematic code from earlier:
\begin{minted}{rust}
v.get_mut(v[0]);
\end{minted}

\noindent We can appease the borrow checker by simply binding the argument to a variable:

\begin{minted}{rust}
let x = v[0];
v.get_mut(x);
\end{minted}

\noindent As a more complex example, let's say we want to build an index over a document that maps each token to a list of occurrences. Before building the index, we apply preprocessing to filter unwanted tokens.

\begin{minted}{rust}
struct Index {
  tokens: Vec<String>,
  index: HashMap<String, Vec<usize>>
}

impl Index {
  fn nonempty_tokens(&self) -> Vec<&String> {
    // Get a vector of *pointers* to strings in self.tokens
    self.tokens.iter().filter(|s| s != &"").collect()
  }

   fn build_index(&mut self) {
    // Mutate self.index while iterating over vector of string pointers
    for (idx, token) in self.nonempty_tokens().into_iter().enumerate() {
      let entry = self.index.entry(token.clone()).or_insert(vec![]);
      entry.push(idx);
    }
  }
}
\end{minted}

\noindent This program fails to compile with the error:

\begin{Verbatim}
error: cannot borrow `self.index` as mutable because it is also borrowed as immutable
14 |     for (idx, token) in self.nonempty_tokens().iter().enumerate() {
   |                         -----------------------------------------
   |                         immutable borrow occurs here
15 |       let entry = self.index.entry((*token).clone()).or_insert(vec![]);
   |                   ^^^^^^^^^^^^^^^^^^^^^^^^^^^^^^^^^^^^^ mutable borrow occurs here
\end{Verbatim}

\noindent The issue is that \rusti{nonempty_tokens} returns an immutable reference to values within the \rusti{tokens} vector. However, the borrow checker reasons conservatively across function boundaries, so an immutable borrow to any field is considered an immutable borrow on the entire struct. Hence, when attempting to mutably update \rusti{self.index} in \rusti{build_index}, the compiler complains of a mutable/immutable conflict.

This kind of pattern presents a challenging design problem. One workaround is to inline the \rusti{nonempty_tokens} function. For example, if we write:

\begin{minted}{rust}
let nonempty_tokens = self.tokens.iter().filter(|s| s != &"").collect();
for (idx, token) in nonempty_tokens.into_iter().enumerate() { .. }
\end{minted}

\noindent This program will pass the borrow checker, because it's able to reason locally about the independence of the \rusti{tokens} and \rusti{index} borrows. But, this approach reduces the modularity of our program. Another alternative is to push the \rusti{nonempty_tokens} method down onto the \rusti{self.tokens} field.

\begin{minted}{rust}
struct Tokens(Vec<String>);
impl Tokens {
  fn nonempty(&self) -> Vec<&String> { self.0.iter().filter(..).collect() }
}

struct Index { tokens: Tokens, .. }
impl Index {
  fn build_index(&mut self) {
    for (idx, token) in self.tokens.nonempty().into_iter().enumerate() { .. }
  }
}
\end{minted}

\noindent The high-level point is that these tricks are not obvious --- they are a form of tacit knowledge that ought to be explicitly taught. Moreover, to understand why the tricks work and when to apply them, a programmer must understand the internals of the borrow checker. To answer "is my code sound?", a programmer needs to only identify a disparity between the high-level rules and a black-box incomplete analyzer. But to actually satisfy the borrow checker (without unsafe-based back-doors), a black-box is no longer sufficient.

Consequently, we need to teach Rust programmers about the actual implementation of the borrow checker, not just the rules. Herein lies an interesting research question: is a more complex analysis (i.e. more complete) on balance easier or harder for programmers to use? In theory, a more complex/complete analysis means fewer situations to wrestle with the analyzer, but a greater challenge when those situations arise. Interestingly, the Rust developers claimed the introduction of non-lexical lifetimes was easier to understand, despite the increase in complexity. Quoting the feature proposal\,\cite{nllrfc}:

\begin{quote}
    Part of the reason that Rust currently uses lexical scopes to determine lifetimes is that it was thought that they would be simpler for users to reason about. Time and experience have not borne this hypothesis out: for many users, the fact that borrows are "artificially" extended to the end of the block is more surprising than not. Furthermore, most users have a pretty intuitive understanding of control flow. [\ldots]

    \

    The primary [drawback] is that the rules for the system become more complex. However, this permits us to accept a larger number of programs, and so we expect that using Rust will feel simpler. Moreover, experience has shown that --- for many users --- the current scheme of tying reference lifetimes to lexical scoping is confusing and surprising.
\end{quote}

\subsection{Is there a function that can bypass the borrow checker?}

Prior work on barriers to adoption for Rust has analyzed dozens of blog posts and forum threads documenting programmers' experiences with learning the language\,\cite{zeng2019identifying}. One common interaction had the form:


\begin{itemize}
    \item \textbf{Newbie}: I can't figure out how to do X! The borrow checker rejects my code, and I can't find any answer about X online.
    \item \textbf{Expert}: Oh, that's because you can't technically do X without \rusti{unsafe}. Did you know about library function Y? It does exactly what you need.
    \item \textbf{Newbie}: Wow! That's perfect. If only I had known...
\end{itemize}

\noindent For example, X = mutably referencing disjoint parts of an array, and Y = \rusti{Vec::split_at_mut}. Or X = temporarily taking ownership of a value of type \rusti{&mut T}, and Y = \rusti{take_mut::take}.

The core issue here is discoverability. In many of the interactions we observed, novices weren't aware of the sound/incomplete distinction, so they didn't think to go looking for a helper function. Even if they did go searching, finding the right function poses a big challenge. For example, if you wanted to find \rusti{Vec::split_at_mut} but didn't know the method name, the Rust documentation for \rusti{Vec} lists 182 methods. Good luck!

I've had echoes of this experience when learning to write proofs in Lean. If I'm writing a proof about natural numbers and my tactics fail, I'll start hunting for a relevant theorem. It's an excruciating experience to carefully scan the hundreds of theorems in the standard library, until I find just the right \verb|nat.pow_lt_pow_of_lt_right| or whatever.

One basic strategy is to create textual resources and automated lints that help identify when users encounter these issues, and point to possible fixes. For example, Esteban K{\"u}ber is the unsung hero of Rust who works on the compiler's error messages. After publication of the Rust barriers to adoption paper\,\cite{zeng2019identifying}, he implemented a help message for rustc that detects the specific pattern of disjoint mutable indexing of an array, and points out \rusti{Vec::split_at_mut} as a possible solution. Here's another research challenge: can we save Esteban the time and automatically generate these lints?

\subsection{Do I need to write that unsafe code myself?}

When all else fails, it's time to program like it's 1972: with \rusti{unsafe} and raw pointers. While uncommon in pure Rust code, these situations do arise. Consider the \rusti{Index} example from above. The hash map used owned strings as keys, which requires an a new heap allocation per key. A more efficient strategy would use to use string pointers as keys, something like:
\begin{minted}{rust}
// 'a is a lifetime variable, NOT a type variable
// 'a is a lower bound on the lifetime of the index string pointers
struct Index<'a> {
  tokens: Vec<String>,
  index: HashMap<&'a str, Vec<usize>>
}
\end{minted}

\noindent This example is a \textit{self-referential struct}, i.e. one field contains pointers to another field. In general, this pattern is unsound. If the \rusti{tokens} field is moved or dropped, the \rusti{&'a str} pointers become invalid. The Rust community has developed user-level abstractions that help with these patterns, e.g. the \rusti{Pin} API that ensures an object will never be moved. However, when designed at the user-level, the compiler isn't aware of these properties, and so \rusti{Pin} must be used in conjunction with raw pointers.

Beyond the obvious safety concerns of dealing with raw pointers (see \cite{qin2020understanding} for gory details), the usability challenge is that unsafe code is an entirely new sub-language to learn. Unsafe programming includes new data structures (\rusti{Pin}, \rusti{UnsafeCell}, \rusti{NonNull}), pointer manipulation APIs (\rusti{std::ptr}, \rusti{std::mem}), and general best practices to avoid self-inflicted foot injuries. Conversely, some of the implicit affordances of Rust's type system fall away --- pointers are no longer automatically dereferenced, for example.

Another concern is whether novices will put in the effort to identify situations where \rusti{unsafe} is necessary. Several people have asked me over the years, ``if Rust lets people write \rusti{unsafe}, won't everyone just get lazy and end up being unsafe?'' This concern is similar to gradually-typed languages like Typescript, where it's just \textit{so easy} to banish a type error with a little ``\rusti{any}'' annotation. For Rust, I'm optimistic that unsafe is sufficiently hard to use and discouraged by the community, such that a novice won't reach for it unless they absolutely need to. Nonetheless, when the time comes for unsafe, Rust users face an uphill battle in learning an unfamiliar language.










\section{To Ownership and Beyond}

In summary, a programmer trying to understand ownership in Rust must learn the following:

\begin{enumerate}
    \item The ownership rules, separated from how they're enforced by the borrow checker.
    \item The distinction between soundness and completeness.
    \item How the borrow checker actually works, and in what situations it's likely to fail.
    \item That functions may exist to make the unsafe into safe, and how to find them.
    \item How to identify when unsafe is absolutely necessary.
\end{enumerate}

\noindent I think just studying these issues, specifically in the context of Rust and its borrow checker, would be a wonderful part of the research agenda for the HATRA community. After all, Rust is the single most widely-used functional programming language \textit{ever}. More than Haskell, Scala, OCaml and Elm combined! The Rust community has hundreds of thousands of people to be both studied and supported in our research.

But beyond Rust and beyond ownership, I think the challenges described here can provide a useful framework for investigating the usability of other type systems. Whether it's dependent types, effect systems, or anything else, we will always have (hopefully) sound and (definitely) incomplete static analyses. Programmers using these systems will need to distinguish between the kind of type errors they receive, and learn to use or avoid type system back-doors.

Solutions to this problem will necessarily come from many domains. Here's a few more research problems to think about:
\begin{itemize}
    \item \textbf{Notional machines:} Researchers in CS education have developed a theory of notional machines, or how students build a mental model of a language's operational semantics\,\cite{sorva2007notional}. What would a notional machine for the borrow checker look like? Perhaps we could present formulations of type systems that look more like runtime semantics\,\cite{kuan2007rewriting}.

    \item \textbf{Visualizations:} type systems like Rust's infer a significant amount of static information, like types and lifetimes. Tools for visualizing this information can help programmers build a mental model of how the type checker works, and reason about why individual errors occur. Displaying this information in a succinct, non-intrusive, yet informative way is still an open problem.

    \item \textbf{Error messages:} besides visualization, the other kind of human-compiler interface is an error message. Errors are responsible for helping programmers localize and interpret syntax/type bugs. Prior work in improving errors has focused on solving the programmer's immediate problem\,\cite{becker2019compiler}. But can error messages also be used pedagogically, to teach the programmer how to avoid a class of type errors in the future?
\end{itemize}

\bibliographystyle{ACM-Reference-Format}
\bibliography{sample-base}

\end{document}